\documentstyle[seceq,epsf]{ptptex}
\begin{document}
\notypesetlogo  

\markboth{
K. Tomita
}{
Anisotropy of the Hubble Constant in a Cosmological Model
}

\title{\Large \bf Anisotropy of the Hubble Constant in a Cosmological
\\ Model with a Local Void on Scales of $\sim 200$ Mpc}
 \author{Kenji {\sc Tomita}} 
\inst{Yukawa Institute for Theoretical Physics, Kyoto University,
        Kyoto 606-8502}

\abst{%
A spherical cosmological model with a local void on scales of $\sim
200$ Mpc and with an inhomogeneous Hubble constant was proposed in
recent two  papers. This model explains consistently the observed properties 
of the cosmic bulk flow, the accelerating behavior of type Ia
supernovae and the CMB dipole anisotropy without invoking a
cosmological constant. As we are in a position deviated from 
the center in the model, the anisotropy of the Hubble constant 
appears owing to the directional difference 
of the distance from the observer to the boundary of the void region. 
It is found that the anisotropy is maximally about 6 $\%$ of the 
constant in the region 
of $200 \sim 500$ Mpc from us. This inhomogeneity and anisotropy of the 
Hubble constant are not so large as to be inconsistent with the present
observation. The detection of this anisotropy in the future will
be useful to clarify the implication of the inhomogeneity of the
Hubble constant. 
}  

\maketitle

\section{Introduction}
\label{sec:level1}

A spherical cosmological model with a local void on scales of 
$\sim 200$ Mpc was proposed in our recent
papers,\cite{toma}\tocite{tomb} (cited as Paper 1
and Paper 2, respectively), to explain consistently the 
observed properties of the cosmic bulk flow, the accelerating behavior 
of type Ia supernovae (SNIa) and the CMB dipole anisotropy, without
invoking a cosmological constant. In this model the Hubble constant is
inhomogeneous. Here we will derive the anisotropy of the constant 
associated with the 
inhomogeneity and discuss its consistency with the present observed 
values of the Hubble constant. 
 
The deviation of local expansion rates $H_{\rm L}$ from the global
Hubble constant $H_0$ due to small-scale perturbations and large-scale 
structures of the universe has so far been studied by many people from 
various viewpoints. Here, let us briefly review the long history of
these studies. First, Turner et al.\cite{tco} derived the probability 
distribution 
function of $H_0$ with given $H_{\rm L}$ using numerical simulations, and
Suto et al.\cite{sut} and Nakamura and Suto\cite{nakam} obtained it 
analytically in a simple
model of inhomogeneities with a local low-density region. It was found 
as a result that the probability of a large deviation such as $(H_{\rm
L} - H_0)/H_{\rm L} > 0.4$ is very small for inhomogeneities with the
radii $\sim 100 h^{-1}$ Mpc, where $H_0 = 100 h$ km s$^{-1}$
Mpc$^{-1}$. In the linear analyses using power spectra in the CDM
models, moreover, Shi and Turner\cite{shi} and Wang et al.\cite{wang} 
showed that the cosmic
and sample variances of $(H_{\rm L} - H_0)/H_0$ are $\sim 0.4 \%$ for
the samples extending to $\sim 100 h^{-1}$ Mpc.

Second, the behavior of CMB dipole anisotropy was discussed to derive
the constraints upon structures with inhomogeneous expansion rates by
Nakao et al.\cite{nakao}, Tomita{\cite{tom95}\tocite{tom96} \cite{toma}},
 \ Shi et al.\cite{swd}, Shi and Turner\cite{shi},  Humphreys et 
al.\cite{hump}, and   Wang et al.\cite{wang}, where the anisotropy was 
assumed to be measured by
an off-center observer in spherical models with an inner low-density
region enclosed by an outer homogeneous region. When we consider only
the redshift arising from the observer's peculiar motion relative to
the CMB rest frame, the dipole component is found to be unacceptably
large for the radius of the low-density region $\sim 200 h^{-1}$
Mpc. The present author, however, showed{\cite{toma} \cite{tom96}}
by solving exactly the null-geodesic equations from the
off-center observer to the last-scattering surface, that in the spherical
models, not only the redshift from the observer's relative motion, but 
also the gravitational redshift are effective, and their main terms are
cancelled out, so that the total dipole component is small compared
with the component only from the relative motion, and reduces to about 
$1/10$ of the latter component, as long as he is near the center.
 Because of this special situation, spherical
structures on the scale $\sim 200 h^{-1}$ with $H_0/H_{\rm L} \sim 0.8$
are not constrained by the observed value of CMB dipole
anisotropy. 

Third, the models with inhomogeneous expansion rates have been studied 
 in connection with the spatial distribution and the $[m, z]$ relation 
of type I supernovae (SNIa) by Wu et al.,\cite{wu95} Wu et al.,\cite{wu96}
 Zehavi et al.\cite{zehav98} and Tomita.\cite{tomb}  Zehavi et 
al.\cite{zehav98} investigated the
local inhomogeneity on the scale of $\sim 60 h^{-1}$ Mpc 
around the Local Group as a model of the {\it Hubble bubble}. However, it
 contradicts with the observational result of Giovanelli et 
al.\cite{giov99} which
shows that the Hubble flow is uniform in the region within $100
h^{-1}$ Mpc.
 
 Independently of their model, the present author proposed another
spherical model (the {\it cosmological void model}), which consists of 
the inner homogeneous region (on the scale of $\sim 200 h^{-1}$ Mpc) 
and the outer homogeneous region with different Hubble constants 
(Paper 1 and Paper 2). 
It was introduced to explain the puzzling situation in
the cosmic bulk flow on the scale of $\sim 150 h^{-1}$ Mpc that was
observed by Hudson et al.\cite{hud99} and Willick.\cite{will99}
Their observational results are not compatible with the other observations 
(Dale et al.,\cite{dale99} Giovanelli et al.,\cite{giov98} and 
Riess et al.\cite{riess97}) in homogeneous
cosmological models, as was discussed by many people in the Workshop
of Cosmic Flows (S. Courteau et al. 1999), but they may be compatible 
with them in our inhomogeneous model (Tomita\cite{tomc}). This is 
because (a) in the inner homogeneous
region there is no peculiar motion between comoving observers and
clusters, (b) the off-center observers find a systematic motion of
comoving clusters in the inner region relative to the global expansion, 
and (c) the total CMB dipole anisotropy measured by the off-center
observers is small, and so the off-center observers in the inner
region feel as if they were in the CMB rest frame, in spite of their 
motions relative to the global expansion. Thus our inhomogeneous 
model may explain the puzzling situation in the bulk flow, though the
probability of its realization is very small.

It was also shown that this model can explain the accelerating behavior of
high-redshift supernovae, without a 
cosmological constant. It does not contradict with Giovanelli et 
al.'s result, because the diameter of the inner region is much 
larger than the size of their observed region $100 h^{-1}$ Mpc.

The local values of the Hubble constant have recently been measured 
by the HST Key Project (Sakai et al\cite{sakai00}), the High-z 
Supernova Search Team,\cite{jha99} and Tutui et al.\cite{tutui00} 
using the common calibrations due to
Cepheid stars. The measurement in the first
group using the multi-wavelength Tully-Fisher relation is limited to 
the region within $100 h^{-1}$ Mpc and the obtained median value is
71 km s$^{-1}$ Mpc$^{-1}$. The unit km s$^{-1}$ Mpc$^{-1}$ is omitted 
in the following for simplicity.  The measurements in the 
second group were done in the region reaching $400 h^{-1}$ Mpc, and
the median value is 64. Those in the third group were performed using
the CO Tully-Fisher relation (Sofue et al.{\cite{sofue96} \cite{sofue99}}) 
in the region between $100$ and $400 h^{-1}$ Mpc and the median value is 
61.  All of these values have errors of the order of $\pm 10$, but  
it seems that these locally measured values of the Hubble constant
 have the tendency to decrease with the increase of the distance.  

In view of the above studies we derive in this paper the effective 
Hubble constant (corresponding to the observed constant) in the above 
inhomogeneous models with a local void 
and study the anisotropy which will be observed by an off-center 
observer. The detection of this anisotropy will be useful to
consider the implication of observed inhomogeneity of the Hubble constant.

\section{Effective Hubble constant}
\label{sec:level2}

Various spherical inhomogeneous models (i.e. single-shell models, 
multi-shell models and models with a self-similar region) were considered
in Paper 1. Here we treat only the simplest model consisting of the inner
low-density homogeneous region (V$^{\rm I}$) and the outer high-density 
homogeneous region (V$^{\rm II}$) connected with a single shell.
The density parameters and Hubble constants in V$^{\rm I}$ and V$^{\rm
II}$ are denoted as $(\Omega_0^{\rm I}, H_0^{\rm I})$ and
$(\Omega_0^{\rm II}, H_0^{\rm II})$, respectively, where $H_0^l = 100
h^l$ \ ($l =$ I or II). Here we assume 
$\Omega_0^{\rm I} < \Omega_0^{\rm II}$ and $H_0^{\rm I} > H_0^{\rm
II}$. The typical Hubble constant is $H_0^{\rm I} = 71$ and $H_0^{\rm
II} = 57 \ (= 0.8 \times H_0^{\rm I})$. The radius of the shell and the
distance from the center C to our observer 
are assumed to be $200 (h^{\rm I})^{-1}$ Mpc and $40 (h^{\rm I})^{-1}$ 
Mpc, as in previous papers. 

In this paper we consider the local behavior of light rays (reaching
the observer) in the near region within 500 Mpc around O, and so the 
distance is expressed using the lowest-order terms of the expansion
with respect to $z (\le 0.15)$. In constrast with  the 
description in previous papers, the paths are expressed using polar 
coordinates $(r,\phi, \theta)$ with the origin O, as in Fig.1.

\begin{figure}
\epsfxsize = 8cm
\centerline{\epsfbox{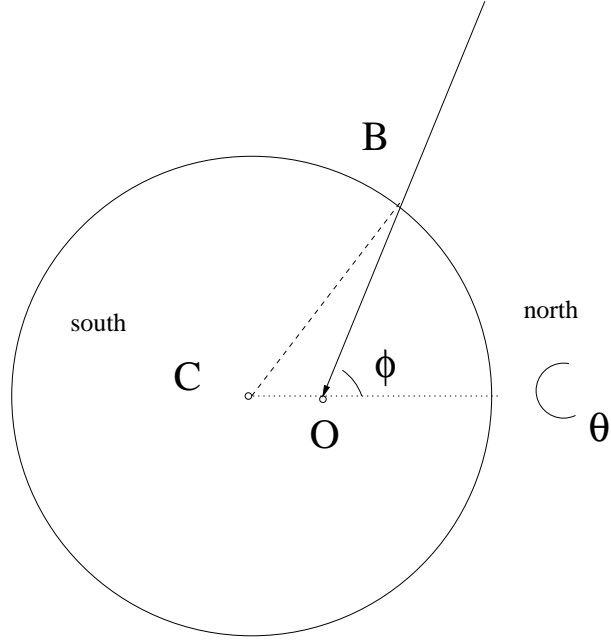}}
\caption{A model with a single shell.  \label{fig1}}
\end{figure}

The center C has the coordinates $r = r_c$ and $\phi = \pi$. In these
coordinates the line-element in V$^{\rm I}$ and V$^{\rm II}$ are
expressed as
\begin{equation}
  \label{eq:m1}
 ds^2 = -c^2 (dt^l)^2  + [a^l (t^l)]^2 \
 \Big\{(d r^l)^2 + [\sinh r^l]^2 d\Omega^2 \Big\},
\end{equation}
where $l =$ I or II and $d\Omega^2 = d\phi^2 + \sin^2 \phi \ d
\theta^2$. The shell is given by $r^l = {r_b}^l(\phi)$, which depends on
time $t^l$, but its time dependence is neglected in the calculation of 
light paths, because $(a^l d {r_b}^l/d t^l)/c \ (\sim 10^{-2})$ is very small 
and the time difference for rays with $\phi = 0$ and $\pi$ is also very 
small.
 
In V$^{\rm I}$ all rays reaching O are radial and straight, because of
homogeneity in V$^{\rm I}$, and we have the relation 
\begin{equation}
  \label{eq:m2}
{1 \over 1+z^{\rm I}}\ = \ {a^{\rm I}(t^{\rm I}) \over a^{\rm I}(t_0^{\rm
I})} \ = \ 1 + \Big({\dot{a}^{\rm I} \over a^{\rm I}}\Big)_0 (t^{\rm I} -
t^{\rm I}_0) 
\end{equation}
or 
\begin{equation}
  \label{eq:m3}
c (t^{\rm I}_0 - t^{\rm I}) = z^{\rm I} c/ H_0^{\rm I}
\end{equation}
between the origin and the boundary, where $H_0^{\rm I} \equiv 
(\dot{a}^{\rm I}/ a^{\rm I})_0$. In the above equations we
neglected the terms of $O(t^{\rm I} - t^{\rm I}_0)^2$.  
Then we have at the boundary
\begin{equation}
  \label{eq:m4}
c (t^{\rm I}_0 - t^{\rm I}_1) = z^{\rm I}_1 c/ H_0^{\rm I}.
\end{equation}

The direction of light rays changes at the boundary in general, but the
amplitude of their change is small and of the order of $\sinh r - r
\approx r^3 \approx (z^{\rm II})^3$, which is neglected in the present
approximation. Accordingly the light rays are regarded as straight, also
in V$^{\rm II}$, and so in V$^{\rm II}$ we have the relation
\begin{equation}
  \label{eq:m5} 
{1+z^{\rm II} \over 1+z^{\rm II}_1} = {a^{\rm II}(t^{\rm II}_1)
\over a^{\rm II}(t^{\rm II})} = 1 + \Big({\dot{a}^{\rm II} \over a^{\rm
II}}\Big)_1 (t^{\rm II}_1 - t^{\rm II})
\end{equation}
or
\begin{equation}
  \label{eq:m6} 
c (t^{\rm II}_1 - t^{\rm II}) = (z^{\rm II} - z^{\rm II}_1) \ c/H_1^{\rm
II},
\end{equation}
where we neglected similarly the terms of $O(t^{\rm II} - t^{\rm II}_1)^2$.
Because $H_1^{\rm II} - H_0^{\rm II} \approx O(z^{\rm I}_1)$, we can
use $H_0^{\rm II}$ in place of $H_1^{\rm II}$ in the present
approximation, and so we obtain
\begin{equation}
  \label{eq:m7} 
c (t^{\rm II}_1 - t^{\rm II}) = (z^{\rm II} - z^{\rm II}_1) \ c/H_0^{\rm
II}.
\end{equation}
At the boundary the equality \ $z^{\rm I}_1 =
z^{\rm II}_1$ \ holds due to the junction condition that the lapses of 
times $t^{\rm I}$ and $t^{\rm II}$ are continuous (cf. Paper I). In the
following we put $z^{\rm
I}_1 (= z^{\rm II}_1), \ z^{\rm I}, \ z^{\rm II}$ as $z_1, \ z, \ z$ for
simplicity. The $\phi$ dependence of the boundary is shown later.

Accordingly the distance $D$ between the observer and the source with
redshift $z$ is
\begin{equation}
  \label{eq:m8} 
D \ (= c(t^{\rm I}_0 - t^{\rm I})) = z \ c/H_0^{\rm I} \quad {\rm for}\  z
\leq z_1(\phi)
\end{equation}
and 
\begin{eqnarray}
  \label{eq:m9}
&D&(= c(t^{\rm I}_0 - t^{\rm I}_1) + c(t^{\rm II}_1 - t^{\rm II}))
= z_1 \ c/H_0^{\rm I} \cr
&&+ (z-z_1) \ c/H_0^{\rm II} \quad {\rm for} \ z > z_1(\phi).
\end{eqnarray}
At the boundary we have
\begin{equation}
  \label{eq:m10}
D_1 (\phi) = z_1 (\phi) \ c/H_0^{\rm I}.
\end{equation}

Here we define the effective Hubble constant $H_0^{\rm eff}$, which
corresponds to the observed Hubble constant, by
\begin{equation}
  \label{eq:m11}
D = z \ c/H_0^{\rm eff}.
\end{equation}
Then $H_0^{\rm eff} = H_0^{\rm I}$ for $z < z_1$, and
\begin{equation}
  \label{eq:m12}
{H_0^{\rm I} \over H_0^{\rm eff}} = {H_0^{\rm I} \over H_0^{\rm II}} -
\Big({H_0^{\rm I} \over H_0^{\rm II}} - 1 \Big) {z_1 \over z}
\end{equation}
for $z \geq z_1$. If we eliminate $z$ using Eqs. (\ref{eq:m9}) and
(\ref{eq:m10}), we obtain 
\begin{equation}
  \label{eq:m13}
H_0^{\rm eff} = H_0^{\rm II} + (H_0^{\rm I} - H_0^{\rm II})
D_1(\phi)/D
\end{equation}
for $z \geq z_1$.

Next the functional form of $D_1(\phi)$ is given. In terms of the distance
$D_b$ (between C and a point B on the boundary) and the distance 
$D_o$ (between C and O), we obtain
\begin{equation}
  \label{eq:m14}
D_1(\phi) = \Big[ D_b^2 - D_o^2 \sin^2 \phi \Big]^{1/2} - D_o \cos \phi
\end{equation}
from the geometrical analysis in $\Delta$COB in Fig.1.
If $D < D_b - D_o$ or $D > D_b + D_o$, we have $D <$ or $> D_1(\phi)$
for all $\phi$. If $D_b + D_o > D> D_b - D_o$, we have $D = D_1
(\phi)$ for $\phi = \phi_1$ specified by
\begin{equation}
  \label{eq:m15}
\mu_1 \equiv \cos \phi_1 \equiv {D_c^2 - D^2 \over 2 D_o D},
\end{equation}
where $D_c \equiv (D_b^2 - D_o^2)^{1/2}.$

\begin{figure}
\epsfxsize = 8cm
\centerline{\epsfbox{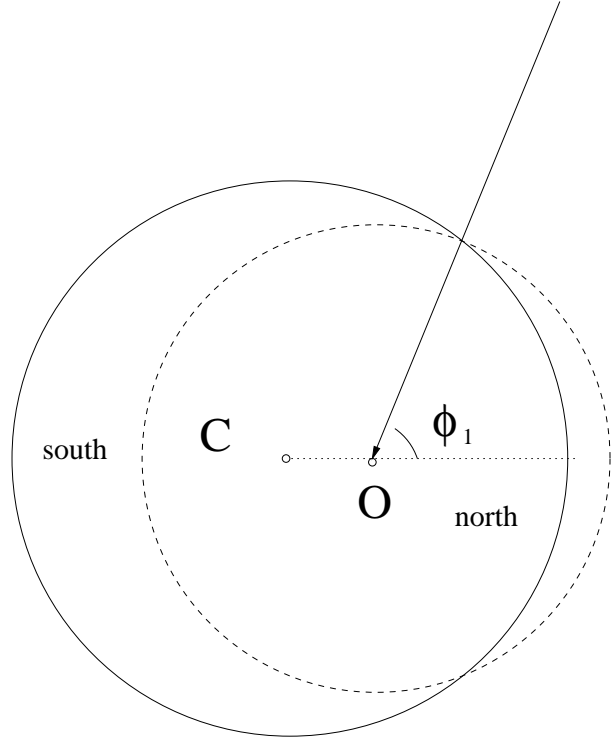}}
\caption{A surface with $D =$ const. in the case of 
$D_c + D_o > D > D_c - D_o$.   \label{fig2}}
\end{figure}

The expressions of the  effective Hubble constant are given as follows 
\ (cf. Fig.2).

 \noindent (1) For $D \leq D_b - D_o$,
\begin{equation}
  \label{eq:m16}
H_0^{\rm eff} = H_0^{\rm I}.
\end{equation}

 \noindent (2) For $D_b + D_o > D> D_b - D_o$,
\begin{equation}
  \label{eq:m17}
H_0^{\rm eff} = H_0^{\rm I} \quad {\rm for} \ \phi \geq \phi_1
\end{equation}
and otherwise $H_0^{\rm eff}$ is given by Eq. (\ref{eq:m13}).

\noindent (3) For $D \geq D_b + D_o, \ H_0^{\rm eff}$ is given by
Eq. (\ref{eq:m13}) similarly. 

\noindent In the case of (2) and (3), $H_0^{\rm
eff}$ has the minimum and maximum at $\phi = 0$ and $\pi$,
respectively.

Now let us consider the angular average of the effective Hubble constant 
($\langle H_0^{\rm eff}\rangle$) for (2) and (3). It is defined by
\begin{equation}
  \label{eq:m18}
\langle H_0^{\rm eff}\rangle = H_0^{\rm II} + {H_0^{\rm I} - H_0^{\rm
II} \over D} I,
\end{equation}
where
\begin{equation}
  \label{eq:m19}
I \equiv \int^{\phi_b}_{\phi_a} {\rm Min} [D, D_1(\phi)] d \cos \phi / 
\int^{\phi_b}_{\phi_a} d \cos \phi
\end{equation}
for the average interval $[\phi_a, \phi_b]$. 
As examples we consider the whole-sky average ($\phi_a = 0, \phi_b = \pi
$), the northern-sky average ($\phi_a = 0, \phi_b = \pi/2$),
and the southern-sky average ($\phi_a = \pi/2, \phi_b = \pi$).
Here the north is taken to be in the direction of \ C $\rightarrow$ O.
The integrals ($I$) corresponding to their averages are denoted as
$I_{\rm A} (i), I_{\rm N} (i)$ and $I_{\rm S} (i)$, respectively, in 
which $i = 2$ and $3$ for the above cases (2) and (3),
respectively. The expressions of these integrals are given in Appendix A.

The maximum and minimum values of $H_0^{\rm eff}$ \ (for $\phi = \pi$ and
$\phi = 0$) and the values $\langle H_0^{\rm eff}\rangle_{\rm i}$ for whole-,
northern- and southern-sky averages (i $=$ A, N and S) were calculated in a
typical example of $D_o = 40 (h^{\rm I})^{-1}$ Mpc, \ $D_b = 5 D_o, \ H_0^{\rm
I} = 71$ and $H_0^{\rm II} = 57$, which were assumed in
our previous papers. Their behaviors are shown in Fig. 3.

\begin{figure}
\epsfxsize =10cm
\centerline{\epsfbox{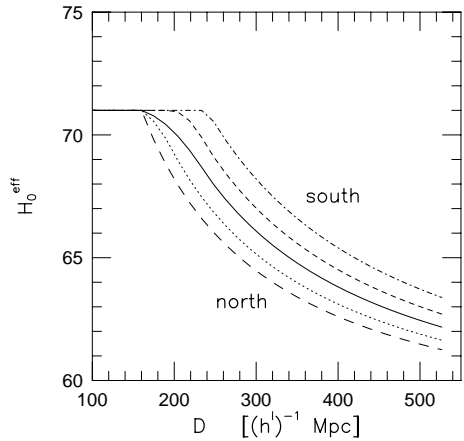}}
\caption{A diagram of the effective Hubble constant
$H_0^{\rm eff}$ (km s$^{-1}$ Mpc$^{-1}$) and the distance $D$
(Mpc).  The lines show the
maximum, the southern-sky average, the whole-sky average, the northern-sky
average, and the minimum in the order from top to bottom.  \label{fig3}}
\end{figure}

As a result it is found that 

\noindent (1) The whole-sky average $\langle H_0^{\rm eff}\rangle_{\rm
A}$ is constant in $D < D_b - D_o$, and decreases gradually from $71$
\ (for $D = D_b -D_o$) to $62$ (for $D \sim 500 (h^{\rm I})^{-1}$
Mpc),

\noindent (2) The difference between $\langle H_0^{\rm eff}\rangle_{\rm
N}$ and $\langle H_0^{\rm eff}\rangle_{\rm S}$ is about $(2.5 \sim
1.3)$ for $D = (160 \sim 500) (h^{\rm I})^{-1}$ Mpc, 

\noindent (3) The difference between the maximum and minimum values of 
$H_0^{\rm eff}$ is about $(4.0 \sim 2.5)$ for $D = (160 \sim 500)
(h^{\rm I})^{-1}$ Mpc. 

\noindent These behaviors of $H_0^{\rm eff}$ seem to be consistent
with the dispersive but decreasing tendency of the observed Hubble constant.

\section{Concluding remarks}

If there is a spherical inhomogeneity in the Hubble constant, the
anisotopy in it and the bulk flow relative to the global expansion
necessarily occur, as long as our observer is not in the center.
If this bulk flow is the one found by Hudson et al. and Willick, the
corresponding anisotropy should be detected in the same direction,
by the observations of nearby SNIa and galaxies (through CO
Tully-Fisher method). Then the north is in the direction of the 
cosmic bulk flow, that is, $l = 260 \pm 15^\circ, \ b = -1 \pm
12^\circ$ (Hudson et al. 1999), or $l = 266^\circ, \ b =
19^\circ$ with $1\sigma$ error (Willick 1999).

In this paper we assumed the simplest {\it cosmological void
models}, in which the Hubble constant changes abruptly at the
boundary. If we assume smoother models with a self-similar
intermediate region, the change in the effective Hubble constant also
may be somewhat slower.
 
With the present assumption, considering only the lowest-order terms with
respect to $z$, the dependence on the density parameter was
neglected. The nonlinear full treatment with respect to $z$ was given
in Paper 2 and the [distance - $z$] relation was found to be 
consistent with SNIa data.

Here a few comments are added as for our void. It refers to the low-density
region ($V^{\rm I}$) with respect to the total matter density. The galactic
number densities in the two regions V$^{\rm I}$ and V$^{\rm II}$ 
depend on the different 
complicated histories of galaxy formation, and their difference at the 
boundary may be rather small compared to the difference of total densities. 

\appendix
\section{Integrals for the averaging}

The integrals ($I$) for the
whole-sky average ($\phi_a = 0, \phi_b = \pi $) for the cases (2) and
(3) are
\begin{equation}
  \label{eq:m20}
I_{\rm A} (2) = {1 \over 4}\Big[2D (\mu_1 +1) + D_o (J_1 - J_2)\Big],
\end{equation}
where 
\begin{equation}
  \label{eq:m21}
J_1 \equiv \Big[\Big({D_c
\over D_o}\Big)^2 -1\Big]^{1/2} +\Big({D_c\over D_o}\Big)^2 \sin^{-1}
{D_o\over D_c} -1,
\end{equation}
\begin{eqnarray}
  \label{eq:m22}
J_2 &\equiv& \mu_1 \Big[\Big({D_c\over D_o}\Big)^2 - (\mu_1)^2
\Big]^{1/2} \cr
&+& \Big({D_c\over D_o}\Big)^2 \sin^{-1} {\mu_1 D_o\over
D_c} - (\mu_1)^2,
\end{eqnarray}
and

\begin{equation}
  \label{eq:m23}
I_{\rm A} (3) = {1 \over 2} D_o \Big\{\Big[\Big({D_c\over D_o}\Big)^2 
-1\Big]^{1/2} + 
\Big({D_c\over D_o}\Big)^2 \sin^{-1} {D_o\over D_c}\Big\}.
\end{equation}

The integrals $I$ for the northern-sky average ($\phi_a = 0, \phi_b = \pi/2$)
and the southern-sky average ($\phi_a = \pi/2, \phi_b = \pi$) are
expressed as $I_{\rm N} (i)$ and $I_{\rm S} (i)$, respectively, with
$i = 2$ and $3$. Here the north is taken to be in the direction of
\ C $\rightarrow$ O.      

For $D_c > D> D_b -D_o$, 
\begin{equation}
  \label{eq:m24}
I_{\rm N} (2) = {1 \over 2} D_o \Big[J_1 -J_2 \Big] + D \mu_1,
\end{equation}
\begin{equation}
  \label{eq:m25}
I_{\rm S} (2) = D.
\end{equation}

For $D_b - D_o > D > D_c$,
\begin{equation}
  \label{eq:m26}
I_{\rm N} (2) = {1 \over 2} D_o J_1,
\end{equation}
\begin{equation}
  \label{eq:m27}
I_{\rm S} (2) = D (\mu_1 +1) - {1 \over 2} D_o J_2.
\end{equation}

For $D > D_b + D_o$,
\begin{equation}
  \label{eq:m28}
I_{\rm N} (3) = I_{\rm A} (3) - {1 \over 2} D_o,
\end{equation}
\begin{equation}
  \label{eq:m29} 
I_{\rm S} (3) = I_{\rm A} (3) + {1 \over 2} D_o.
\end{equation}

The author is grateful to Prof. T.P. Singh for his careful reading of the 
 manuscript.
 This work was supported by Grant-in Aid for Scientific Research 
(No. 10640266) from the Ministry of Education, Science, Sports and
Culture, Japan.

\def\apj{Astrophys. J.}
\def\apjs{Astrophys. J. Suppl.}
\def\aj{ Astron. J.}
\def\mnras{ Month. Notices R.A.S.}
\def\pasj{Publ. Astron. Soc. Japan}
\def\aap{Astron. Astrophys.}

\end{document}